\documentclass[12pt]{article}

\usepackage{amsthm}
\theoremstyle{plain}

\theoremstyle{definition}

\theoremstyle{proposition}

\theoremstyle{lemma}

\theoremstyle{remark}

\usepackage{amsmath}
\usepackage{amssymb}
\usepackage[dvipdfmx]{graphicx}

\usepackage[bookmarksnumbered=true,colorlinks=true,filecolor=blue,linkcolor=blue,urlcolor=blue,citecolor=red,linktocpage=true]{hyperref}

\makeatletter
 
  \@addtoreset{equation}{section}
 \makeatother

\begin{document}
\setlength{\oddsidemargin}{0cm}
\setlength{\baselineskip}{7mm}

\begin{titlepage}
\begin{flushright}  USTC-ICTS/PCFT-20-03 \end{flushright} 

~~\\

\vspace*{0cm}
    \begin{large}
       \begin{center}
         {Perturbative string theory from Newtonian limit of string geometry theory}
       \end{center}
    \end{large}
\vspace{1cm}

\begin{center}
           Matsuo S{\sc ato}$^{*}$\footnote
           {
e-mail address : msato@hirosaki-u.ac.jp}
and  Yuji S{\sc ugimoto}$^{\dagger,\ddagger}$\footnote
           {
e-mail address : sugimoto@ustc.edu.cn}\\
      \vspace{1cm}
       
         {$^{*}$\it Graduate School of Science and Technology, Hirosaki University\\ 
 Bunkyo-cho 3, Hirosaki, Aomori 036-8561, Japan}\\

{$^{\dagger}$\it Interdisciplinary Center for Theoretical  Study, University of Science and Technology of China, Hefei, Anhui 230026, China}
\\
{$^{\ddagger}$\it Peng Huanwu Center for Fundamental Theory, Hefei, Anhui 230026, China}
                    
\end{center}

\hspace{5cm}

\begin{abstract}
\noindent
String geometry theory is one of the candidates of the non-perturbative formulation of string theory. In \cite{Sato:2017qhj}, the perturbative string theory is reproduced from a string geometry model coupled with a $u(1)$ gauge field on string manifolds. In this paper, we generalize this result and we show that the perturbative string theory is reproduced from any string geometry model by taking a Newtonian limit.

\end{abstract}

\vfill
\end{titlepage}
\vfil\eject

\setcounter{footnote}{0}

\section{Introduction}\label{intro}
\setcounter{equation}{0}
String geometry theory is one of  the candidates of non-perturbative formulation of string theory \cite{Sato:2017qhj}. 
Especially, from a sting geometry model coupled with  a $u(1)$ gauge field\footnote{The action of string geometry theory is not determined as of this moment. On this stage, we should consider various possible actions. Then, we call each action a string geometry model and call the whole formulation string geometry theory.}, one can derive the all-order perturbative scattering amplitudes that possess the super moduli in IIA, IIB and SO(32) I superstring theories, by considering fluctuations around fixed perturbative IIA, IIB and SO(32) I vacuum background, respectively\footnote{Thus, the critical dimension is determined to be 10 perturbatively in string geometry theory. This is a condition for the fixed background in string geometry. This is the same situation with the perturbative string theory, where the critical dimension is also determined as a condition for a string background, although the background is fixed when the theory is formulated in this case. We expect that the critical dimension is determined to be 10 without fixing any background in string geometry theory by performing the BRST quantization of the theory, without moving to the first quantization formalism as in the derivation of the perturbative string theory.}\footnote{A perturbative topological string theory is also derived from topological string geometry theory \cite{Sato:2019cno}. } as in \cite{Sato:2017qhj}.  

Then, a natural question arises as to whether a perturbative string theory can be reproduced if a string geometry model couples with other fields.  Is the $u(1)$ gauge field special? Or, can it be reproduced from any string geometry model?

To answer this question, let us reconsider how the perturbative string theory is reproduced from the string geometry model coupled with the $u(1)$ gauge field. In this case, it is derived from fluctuations around a background including a potential that satisfies the harmonic equation on the flat background. A low-velocity limit is also taken. These facts indicate Newtonian limit\footnote{This was pointed out by H. Kawai.}. Then, we examine all the string geometry models by taking a Newtonian limit in this paper.  

The organization of the paper is as follows. In section 2,  we shortly review the framework of string geometry theory. In section 3, we show that a Newtonian limit of any string geometry model reproduces the perturbative string theory. In section 4, we conclude and discuss our results. In appendix A, we clarify the relation between the results of this paper and of the previous paper in case of coupling with a $u(1)$ gauge field on string manifolds. In appendix B, we display an explicit form of the Lagrangian up to the second order in the fluctuations.

\vspace{1cm}

\section{String geometry theory}
\setcounter{equation}{0}
In this paper, we discuss only the closed bosonic sector of string geometry theory \cite{Sato:2017qhj}. One can generalize the result in this paper to the full string geometry theory in the same way as in \cite{Sato:2017qhj}. The closed bosonic sector is described by a partition function
\begin{align}
Z=\int \mathcal{D}G \mathcal{D}Ce^{-S},
\end{align}
where the action in general is given by
\begin{align}
S=\frac{1}{G_N}\int \mathcal{D} h \mathcal{D} X(\bar{\tau})\mathcal{D} \bar{\tau} 
\sqrt{G} (-R + (\mbox{tensor and higher derivative terms}) ), \label{BosonicAction}
\end{align}
where $G_{N}$ is a constant.
The action consists of a metric $G_{IJ}$ and tensor fields $C$ including scalar fields defined on an infinite dimensional manifold, so-called string manifold. String manifold is constructed by patching open sets in string model space $E$, whose definition is summarized as follows.  First, a global time $\bar{\tau}$ is defined canonically and uniquely on a Riemann surface $\bar{\Sigma}$ by the real part of the integral of an Abelian differential uniquely defined on $\bar{\Sigma}$ \cite{KricheverNovikov1, KricheverNovikov2}.
We restrict $\bar{\Sigma}$ to a $\bar{\tau}$ constant line and obtain $\bar{\Sigma}|_{\bar{\tau}}$. An embedding of $\bar{\Sigma}|_{\bar{\tau}}$ to $\mathbb{R}^{d}$ represents a many-body state of strings in $\mathbb{R}^{d}$, and is parametrized by coordinates $(\bar{h}, X (\bar{\tau}), \bar{\tau})$\footnote{`` $\bar{}$ " represents a representative of the diffeomorphism and Weyl transformations on the worldsheet.} where $\bar{h}$ is a metric on  $\bar{\Sigma}$ and $X (\bar{\tau})$ is a map from  $\bar{\Sigma}|_{\bar{\tau}}$ to $\mathbb{R}^{d}$.  String model space $E$  is defined by the collection of the string states by considering all the  $\bar{\Sigma}$, all the values of $\bar{\tau}$, and all the $X (\bar{\tau})$. How near the two string states is defined by how near the values of $\bar{\tau}$ and how near $X (\bar{\tau})$\footnote{The precise definition of the string topology is given in the section 2 in \cite{Sato:2017qhj}.}. By this definition, arbitrary two string states on a connected  Riemann surface in $E$ are connected continuously. Thus, there is an one-to-one correspondence between a Riemann surface in $\mathbb{R}^{d}$ and a curve  parametrized by $\bar{\tau}$ from $\bar{\tau}=-\infty$ to $\bar{\tau}=\infty$ on $E$. That is, curves that represent asymptotic processes on $E$ reproduce the right moduli space of the Riemann surfaces in $\mathbb{R}^{d}$. Therefore, a string geometry model possesses all-order information of string theory.

We use the Einstein notation for the index $I$, where $I=\{d,(\mu \bar{\sigma}) \}$.
The cotangent space is spanned by $d X^{d} := d \bar{\tau}$ and $d X^{(\mu \bar{\sigma})  }:= d X^{\mu} \left( \bar{\sigma}, \bar{\tau} \right)$ for $\mu=0,1, \dots, d-1$, while $d \bar{h}_{mn}$ with $m,n=\bar{\tau},\bar{\sigma}$ cannot be a part of the basis because $\bar{h}_{mn}$ is treated as a discrete valuable in the string topology. 
The summation over $\bar{\sigma}$ is defined by $\int d\bar{\sigma}  \bar{e} (\bar{\sigma}, \bar{\tau})$, where $\bar{e}:=\sqrt{\bar{h}_{ \bar{\sigma} \bar{\sigma}}}$. This summation is transformed as a scalar under $\bar{\tau} \mapsto \bar{\tau}'(\bar{\tau},  X(\bar{\tau}))$, and invariant under $\bar{\sigma} \mapsto \bar{\sigma}'(\bar{\sigma})$.

From these definitions, we can write down the general form of the metric of the string geometry as follows.
\begin{align}
ds^2 &(\bar{h}, X(\bar{\tau}), \bar{\tau}) 
\nonumber \\
=
&G_{dd} (\bar{h}, X(\bar{\tau}), \bar{\tau}) (d\bar{\tau})^2
+2 d\bar{\tau} \int d\bar{\sigma}  \bar{e} (\bar{\sigma}, \bar{\tau})  \sum_{\mu}  G_{d \; (\mu \bar{\sigma})}(\bar{h}, X(\bar{\tau}), \bar{\tau}) d X^{\mu}(\bar{\sigma}, \bar{\tau}) \nonumber \\
&+\int d\bar{\sigma}   \bar{e} (\bar{\sigma}, \bar{\tau}) \int d\bar{\sigma}' \bar{e} (\bar{\sigma}', \bar{\tau})  \sum_{\mu, \mu'} G_{ \; (\mu \bar{\sigma})  \; (\mu' \bar{\sigma}')}(\bar{h}, X(\bar{\tau}), \bar{\tau}) d X^{\mu}(\bar{\sigma}, \bar{\tau}) d X^{\mu'}(\bar{\sigma}', \bar{\tau}). \nonumber \\
\end{align}
In this way, string geometry theory is a field theory on a loop space. Thus, we can expect that the theory includes non-perturbative effects as string field theory.

\vspace{1cm}

\section{Newtonian limit of string geometry theory}
\setcounter{equation}{0}
In this section, we show that a Newtonian limit of the string geometry model reproduces the perturbative string theory. 

The Newtonian limit of string geometry theory is defined by (0) low-velocity limit under three conditions:
\begin{itemize}
\vspace{-1mm}
\item[(1)] weak gravity condition
\vspace{-3mm}
\item[(2)] stationary condition for the background
\vspace{-3mm}
\item[(3)] tensor-less condition\footnote{Matters (tensors) are zero in this case, whereas they are sources in the Newtonian limit of general relativity in the four dimensions.}.
\vspace{-1mm}
\end{itemize}

First, we consider the Newtonian potential background around a flat background:
\begin{eqnarray}
\bar{ds}^2
&=& 2\lambda \bar{\rho}(\bar{h})  (1+\alpha  \phi(X(\bar{\tau}))) (dX^d)^2  \nonumber \\
&&+\int d\bar{\sigma}   \bar{e} \int d\bar{\sigma}' \bar{e}' (1+\frac{1}{2-D} \alpha \phi(X(\bar{\tau}))) \frac{\bar{e}^3(\bar{\sigma}, \bar{\tau})}{\sqrt{\bar{h}(\bar{\sigma}, \bar{\tau})}} \delta_{(\mu \bar{\sigma}) (\mu' \bar{\sigma}')}
d X^{(\mu \bar{\sigma})} d X^{(\mu' \bar{\sigma}')}, \label{solution}
\end{eqnarray}
where $\lambda$ is a constant and $\bar{\rho}(\bar{h}):=\frac{1}{4 \pi}\int d\bar{\sigma} \sqrt{\bar{h}}\bar{R}_{\bar{h}}$, where $\bar{R}_{\bar{h}}$ is the scalar curvature of $\bar{h}_{ mn}$. The worldsheet metric $\bar{h}_{mn}$ behaves as a constant in the solution \eqref{solution} to the equations of motion of \eqref{BosonicAction} because they are differential equations with respect to $X^{\mu}(\bar{\tau})$ and $\bar{\tau}$. $D$ is a volume of the index $(\mu \bar{\sigma})$, $D:=\int d \bar{\sigma} \bar{e} \delta_{(\mu \bar{\sigma}) (\mu \bar{\sigma})}= 2 \pi d \delta(0)$\footnote{Here we treat $D$ as a finite parameter for a regularization and take $D\to \infty$  in the end.}. 

The equations of motion reduces to the Ricci flat condition,
\begin{equation}
R_{IJ}=0, \label{RicciFlat}
\end{equation}
because the tensor and higher derivative terms are dropped off from the limit (0) and the condition (3). In addition, the Ricci flat condition reduces to a harmonic equation,
\begin{equation}
\int d \bar{\sigma} \bar{e} \frac{d}{d X^{(\mu \bar{\sigma})}} \frac{d}{d X^{(\mu \bar{\sigma})}} \phi(X(\bar{\tau}))=0,
\label{diffeq}
\end{equation}
because the condition (1) implies that $\phi \ll 1$ where $\phi$ remains up to the second order in the action and up to the first order in the equations of motion, and the condition (2) implies that $\frac{d}{d \bar{\tau}} \phi=0$.
\begin{equation}
\phi(X(\bar{\tau}))= {\rm i} \int d\bar{\sigma}  \epsilon_{\mu\nu}X^{\mu}(\bar{\tau}) \partial_{\bar{\sigma}} X^{\nu}(\bar{\tau})
\label{DefPhi}
\end{equation}
is a solution to the harmonic equation,
where \footnote{In the following, we assume that $d$ is an arbitrary even number however, we will show that $d$ has to be $d=26$ in order to be anomaly free for the perturbative string theory.}
\begin{align}
\epsilon_{\mu \nu} = -\epsilon_{\nu\mu}=
\begin{cases}
 1~\text{ for } (\mu,\nu)=(0,1),(2,3),(4,5),...,(d-2,d-1),
\\
0 ~\text{ for others}.
\end{cases}
\end{align}  
We should note that $\bar{h}_{mn}$, $X^{\mu}(\bar{\tau})$ and $\bar{\tau}$ are all independent, and thus $\frac{\partial}{\partial \bar{\tau}}$ is an explicit derivative on functions over the string manifolds, especially, $\frac{\partial}{\partial \bar{\tau}}\bar{h}_{ mn}=0$ and $\frac{\partial}{\partial \bar{\tau}}X^{\mu}(\bar{\tau})=0$. 

The dependence of $\bar{h}_{mn}$ on the background (\ref{solution}) is uniquely determined  by the consistency of the quantum theory of the fluctuations around the background. Namely, a propagator of the fluctuations becomes to be the path-integral of the perturbative string on the flat background as we will see in the remaining part of this paper. The dependence  is uniquely determined if one assumes that the theory has the two-dimensional diffeomorphism times Weyl invariance quantum mechanically.  The symmetry will be broken if one chooses the other dependence. This mechanism corresponds to the fact that a background is restricted to satisfy the equations of motion of the gravity if one supposes the two-dimensional diffeomorphism times Weyl invariance quantum mechanically in the perturbative string theory. 

Let us consider fluctuations around the background \eqref{solution}, $G_{IJ}=\bar{G}_{IJ}+\tilde{G}_{IJ}$. Because of the limit (0) and the condition (3), the action \eqref{BosonicAction} up to the quadratic order is given by
\begin{eqnarray}
S&=&\frac{1}{G_N} \int \mathcal{D} h \mathcal{D} X(\bar{\tau})\mathcal{D} \bar{\tau}  \sqrt{\bar{G}} 
\Bigl(-\bar{R}-(\bar{R}^{IJ}-\frac{1}{2}\bar{G}^{IJ}\bar{R})\tilde{G}_{IJ}
\nonumber \\
&&+\frac{1}{4}\bar{\nabla}_I \tilde{G}_{JK} \bar{\nabla}^I \tilde{G}^{JK}
-\frac{1}{4}\bar{\nabla}_I \tilde{G} \bar{\nabla}^I \tilde{G}
+\frac{1}{2}\bar{\nabla}^I \tilde{G}_{IJ} \bar{\nabla}^J \tilde{G}
-\frac{1}{2}\bar{\nabla}^I \tilde{G}_{IJ} \bar{\nabla}_K \tilde{G}^{JK}
\nonumber \\
&&+\frac{1}{4}\bar{R}
(\tilde{G}_{IJ}\tilde{G}^{IJ}-\frac{1}{2}\tilde{G}^2)
-\frac{1}{2}\bar{R}^{I}_{\;\; J}
\tilde{G}_{IL}\tilde{G}^{JL}
+\frac{1}{2}\bar{R}^{IJ}
\tilde{G}_{IJ}\tilde{G}
-\frac{1}{2}\bar{R}^{IJKL}
\tilde{G}_{IK}\tilde{G}_{JL} \Bigr),
\end{eqnarray}
where $\tilde{G}:=\bar{G}^{IJ}\tilde{G}_{IJ}$. In order to fix the diffeomorphism symmetry, we take the harmonic gauge. If we add the gauge fixing term
\begin{equation}
S_{fix}=\frac{1}{G_N}\int \mathcal{D} h \mathcal{D} X(\bar{\tau})\mathcal{D} \bar{\tau}  \sqrt{\bar{G}} 
\frac{1}{2} \Bigl( \bar{\nabla}^J(\tilde{G}_{IJ}-\frac{1}{2}\bar{G}_{IJ}\tilde{G}) \Bigr)^2,\end{equation}
we obtain
\begin{align}
S'+S_{fix}
&=\frac{1}{G_N} \int \mathcal{D} h \mathcal{D} X(\bar{\tau})\mathcal{D} \bar{\tau}  \sqrt{\bar{G}} 
\Bigl(-\bar{R}-(\bar{R}^{IJ}-\frac{1}{2}\bar{G}^{IJ}\bar{R})\tilde{G}_{IJ}
\nonumber \\
&\qquad
+\frac{1}{4}\bar{\nabla}_I \tilde{G}_{JK} \bar{\nabla}^I \tilde{G}^{JK}
-\frac{1}{8}\bar{\nabla}_I \tilde{G} \bar{\nabla}^I \tilde{G}
+\frac{1}{4}\bar{R}
(\tilde{G}_{IJ}\tilde{G}^{IJ}-\frac{1}{2}\tilde{G}^2)
-\frac{1}{2}\bar{R}^{I}_{\;\; J}
\tilde{G}_{IL}\tilde{G}^{JL}
\nonumber \\
&\qquad
+\frac{1}{2}\bar{R}^{IJ}
\tilde{G}_{IJ}\tilde{G}
-\frac{1}{2}\bar{R}^{IJKL}
\tilde{G}_{IK}\tilde{G}_{JL} \Bigr). \label{fixedaction}
\end{align}
We normalize the fields as $\tilde{H}_{IJ}:=Z_{IJ} \tilde{G}_{IJ}$, where 
$Z_{IJ}:=G_N^{-\frac{1}{2}} 
\bar{G}^{\frac{1}{4}} 
(\bar{\alpha}_I \bar{\alpha}_J)^{-\frac{1}{2}}$.
$\bar{\alpha}_{I}$ represent the background metric as $\bar{G}_{IJ}=\bar{\alpha}_I \delta_{IJ}$, where $\bar{\alpha}_d=2\lambda\bar{\rho}$ and $\bar{\alpha}_{(\mu \bar{\sigma})}= \frac{\bar{e}^3}{\sqrt{\bar{h}}}$.  By using (\ref{solution}),  we obtain
\begin{align}
S'+S_{fix}=  &\int  \mathcal{D} h \mathcal{D} X(\bar{\tau})\mathcal{D} \bar{\tau} \left(L_0 + L_1 + L_{2} \right), \label{fluctuationAction}
\end{align}
where $L_{0,1,2}$ are the terms of order $\mathcal{O}(\tilde{H}_{IJ}^{0,1,2})$. The explicit form is given in Appendix \ref{DerivExpResult}.

In order to take (0) the low-velocity limit,  we perform a derivative expansion of $\tilde{G}_{IJ}$. We perform  (1) weak gravity expansion and the derivative expansion
\begin{subequations}
\begin{align}
\tilde{H}_{IJ} &\to  \frac{1}{\alpha} \tilde{H}_{IJ}  
 \\
\partial_K \tilde{H}_{IJ} &\to   \partial_K\tilde{H}_{IJ}  
 \\
\partial_K\partial_L\tilde{H}_{IJ} &\to  \alpha \partial_K\partial_L\tilde{H}_{IJ}   \\
G_N &\to  \alpha^4 G_N 
\end{align}
\label{derivexp}
\end{subequations}
and take the Newtonian limit
\begin{equation}
\alpha \to 0,
\end{equation}
where $\alpha$ is an arbitrary constant in the solution (\ref{solution}).  We have also redefined $G_N$.
Then, (\ref{fixedaction}) with appropriate boundary conditions reduces to
\begin{equation}
S'+S_{fix} \to \int \mathcal{D} h \mathcal{D} X(\bar{\tau})\mathcal{D} \bar{\tau}  \left( \tilde{L}_0 + \tilde{L}_1+ \tilde{L}_2 \right),
\end{equation}
where
\begin{align}
&\tilde{L}_0
=
\frac{1}{\alpha^2 G_N} \left(  \frac{3D^2-9D+10}{4(D-2)^2} \bar{\alpha}_d^{-2} \bar{\alpha}_i^{-1} \partial^i \phi \partial_i \phi \right),
\\ 
&\tilde{L}_1=
\frac{1}{\alpha \sqrt{G_N}} \biggl\{ -\frac{(D-1)(D-6)}{8(D-2)^2}\alpha^2 \bar{\alpha}_d^{-2} \bar{\alpha}_i^{-1}  \partial^i \phi \partial_i \phi \tilde{H}_{dd}  
+\frac{3(D-1)}{8(D-2)} \alpha^2  \bar{\alpha}_d^{-2} \bar{\alpha}_i^{-1}  \partial^i \phi \partial_i \phi \tilde{H}^j_{j}
 \nonumber \\
&\qquad\qquad\qquad
+ \biggl(
- \frac{D-1}{4(D-2)} \alpha^2 \bar{\alpha}_d^{-1}  \partial_i  \phi \partial_j \phi 
+ \frac{D-1}{4(D-2)} \alpha^2 \bar{\alpha}_d^{-1} \phi \partial_i \partial_j \phi  \biggr) \bar{\alpha}_d^{-1} \bar{\alpha}_i^{-\frac{1}{2}} \bar{\alpha}_j^{-\frac{1}{2}}   \tilde{H}^{ij}
\biggr\},
\end{align}
\begin{align}
&L_{2}=
\bar{\alpha}_d^{-1}
 \tilde{H}_{dd} 
\biggl\{
-\frac{1}{8} \partial_d^2 
-\frac{1}{8}  \bar{\alpha}_i^{-1} \partial^i \partial_i 
-\frac{D^2-7D+6}{32(D-2)^2}  \alpha^2 \bar{\alpha}_d^{-1} \bar{\alpha}_i^{-1} \partial^i \phi \partial_i  \phi 
 \biggr\}
 \tilde{H}_{dd} 
\nonumber \\
 &\quad\quad
+ \bar{\alpha}_d^{-1}
\tilde{H}_{di} 
\biggl\{
-\frac{1}{2} \bar{\alpha}_i^{-1} \delta^{ij}  \partial_d^2
-\frac{1}{2} \bar{\alpha}_k^{-1}   \delta^{ij} \partial^k \partial_k
+\frac{1}{4(D-2)} \bar{\alpha}_i^{-\frac{1}{2}} \alpha_j^{-\frac{1}{2}}  \partial^i \phi \partial^j   
-\frac{1}{2(D-2)}  \bar{\alpha}_i^{-\frac{1}{2}} \alpha_j^{-\frac{1}{2}}  \partial^j \phi  \partial^i 
\nonumber  \\
&\qquad\qquad\qquad\quad
-\frac{D-3}{2(D-2)} \bar{\alpha}_k^{-1}   \partial_k \phi \delta^{ij} \partial_k 
-\frac{D-9}{8(D-2)} \bar{\alpha}_d^{-1} \delta^{ij} \bar{\alpha}_k^{-1} \partial^k \phi \partial_k  \phi 
\nonumber  \\
&\qquad\qquad\qquad\quad
+\frac{D-3}{2(D-2)}\bar{\alpha}_d^{-1} \bar{\alpha}_i^{-\frac{1}{2}}  \bar{\alpha}_j^{-\frac{1}{2}} \partial_i \phi \partial_j  \phi 
\biggr\} 
\tilde{H}_{dj} 
\nonumber  \\
 &\qquad\quad
+\tilde{H}_{dd}
\biggl\{
\frac{1}{4} \bar{\alpha}_d^{-1}  \delta^{ij}   \partial_d^2
+\frac{1}{4} \bar{\alpha}_d^{-1} \bar{\alpha}_k^{-1} \bar{\alpha}_d \bar{\alpha}_j \delta^{ij}  \partial^k \partial_k 
-\frac{3D-5}{8(D-2)} \bar{\alpha}_d^{-1} \bar{\alpha}_k^{-1} \delta^{ij}   \partial_k \phi  \partial^k 
\nonumber  \\
&\qquad\qquad\qquad\quad
-\frac{1}{2} \bar{\alpha}_d^{-1} \bar{\alpha}_i^{-\frac{1}{2}} \bar{\alpha}_i^{-\frac{1}{2}}  \phi \partial_j \phi  \partial_i  \phi  \partial_j \phi    
-\frac{1}{2} \bar{\alpha}_d^{-1} \bar{\alpha}_i^{-\frac{1}{2}} \bar{\alpha}_i^{-\frac{1}{2}} \overset{\leftarrow}{\partial_i} 
\nonumber  \\
&\qquad\qquad\qquad\quad
+\frac{5D^2-27D+38}{16(D-2)^2} \bar{\alpha}_d^{-2}  \bar{\alpha}_j  \bar{\alpha}_k^{-1}   \delta^{ij}  \partial^k \phi \partial_k \phi 
+\frac{3D-7}{8(D-2)}  \bar{\alpha}_d^{-2} \bar{\alpha}_i^{-\frac{1}{2}}  \bar{\alpha}_j^{-\frac{1}{2}} \partial_i \phi \partial_j  \phi 
\biggr\} 
 \tilde{H}_{ij} 
\nonumber \\
&\qquad\quad
+ \tilde{H}_{ij} 
\biggl\{
\frac{1}{8} \bar{\alpha}_d^{-1} \biggl( -2 \delta^{ik}\delta^{jl}  \partial_d^2 + \delta^{ij}\delta^{kl} \partial_d^2 \biggr)
-\frac{1}{8(D-2)}\bar{\alpha}_d^{-1} \bar{\alpha}_m^{-1}  \partial_m \phi \biggl( -  2 \delta^{ik}\delta^{jl}  \partial^m + \delta^{ij}\delta^{kl} \partial^m \biggr)
\nonumber \\
&\qquad\qquad\qquad\quad
+\frac{1}{8}  \bar{\alpha}_m^{-1} \biggl(- 2 \delta^{ik}\delta^{jl} \partial^m  \partial_m +  \delta^{ij}\delta^{kl} \partial^m \partial_m  \biggr)
+\frac{3D^2-9D-2}{16(D-2)^2} \bar{\alpha}_d^{-2} \bar{\alpha}_m^{-1} \partial^m \phi \partial_m  \phi  \delta^{ik}\delta^{jl} 
\nonumber  \\
&\qquad\qquad\qquad\quad
-\frac{3^2-9D-10}{32(D-2)^2}  \bar{\alpha}_d^{-2} \bar{\alpha}_m^{-1} \partial^m \phi \partial_m  \phi  \delta^{ij}\delta^{kl} 
+\frac{D^2-3D-4}{4(D-2)^2} \alpha \bar{\alpha}_d^{-2} \bar{\alpha}_i^{-\frac{1}{2}} \bar{\alpha}_j^{-\frac{1}{2}}  \partial_i \phi \partial_k  \phi \delta^{jl}
\nonumber  \\
&\qquad\qquad\qquad\quad
+\frac{D^2-3D-4}{4(D-2)^2} \bar{\alpha}_d^{-2}  \bar{\alpha}_i^{-\frac{1}{2}} \bar{\alpha}_j^{-\frac{1}{2}}  \partial_i \phi \partial_j \phi  \delta^{kl} 
-\frac{1}{D-2} \bar{\alpha}_d^{-1} \bar{\alpha}_i^{-\frac{1}{2}} \alpha_k^{-\frac{1}{2}} \partial_i \phi \delta^{jl} \partial^k 
\nonumber  \\
&\qquad\qquad\qquad\quad
+\frac{1}{2(D-2)} \bar{\alpha}_d^{-1} \bar{\alpha}_i^{-\frac{1}{2}} \bar{\alpha}_k^{-\frac{1}{2}}  \partial_k  \phi \delta^{jl} \partial_i
+\frac{D^2+3D-14}{4(D-2)^2} \bar{\alpha}_d^{-2} \bar{\alpha}_i^{-\frac{1}{2}} \alpha_k^{-\frac{1}{2}} \partial_i \phi \partial_k \phi \delta^{jl}
\nonumber  \\
&\qquad\qquad\qquad\quad
-\frac{1}{2(D-2)} \bar{\alpha}_d^{-1} \bar{\alpha}_k^{-\frac{1}{2}} \bar{\alpha}_l^{-\frac{1}{2}}  \partial_l  \phi \delta^{ij}  \partial_k
-\frac{1}{2(D-2)} \bar{\alpha}_d^{-1} \bar{\alpha}_i^{-\frac{1}{2}} \bar{\alpha}_j^{-\frac{1}{2}}   \partial_j  \phi \delta^{ij} \partial_i
\biggr\}  \tilde{H}_{kl}  
\end{align}
where we use short notation: we use indices $i$ instead of $(\mu \bar{\sigma})$ and all the indices are summed up.

In the same way as in \cite{Sato:2017qhj}, a part of the action 
\begin{equation}
\int \mathcal{D} h \mathcal{D} X(\bar{\tau})\mathcal{D} \bar{\tau} 
\int_0^{2\pi}d\bar{\sigma} \tilde{H}^{\bot}_{d(\mu \bar{\sigma})} 
H
\tilde{H}^{\bot}_{d(\mu \bar{\sigma})} \label{SecondOrderAction}
\end{equation}
with 
\begin{align}
H
&=
-\frac{1}{2}\frac{1}{2\lambda\bar{\rho}}(\frac{\partial}{\partial \bar{\tau}})^2
-\frac{1}{2}\int_0^{2\pi} d \bar{\sigma} \frac{\sqrt{\bar{h}}}{\bar{e}^2} (\frac{\partial}{\partial X^{\mu}(\bar{\tau})})^2 
+\frac{1}{2}\int_0^{2\pi} d \bar{\sigma} \frac{\sqrt{\bar{h}}}{\bar{e}^2} \partial_{\bar{\sigma}}X^{\mu}(\bar{\tau})\partial_{\bar{\sigma}}X_{\mu}(\bar{\tau}) 
\label{Hamiltonian}
\end{align}
decouples from the other modes, where we have taken $D \to \infty$, and write down the summation and integration explicitly.

The following derivation is the same as in \cite{Sato:2017qhj}. By adding following identity\footnote{The proof is given in \cite{Sato:2017qhj}.}
\begin{eqnarray}
0
&=&\int \mathcal{D} h \mathcal{D} X(\bar{\tau})\mathcal{D} \bar{\tau} 
\int_0^{2\pi}d\bar{\sigma}' 
\tilde{H}^{\bot}_{d(\mu \bar{\sigma}')} 
( \int_0^{2\pi} d \bar{\sigma}
\bar{n}^{\bar{\sigma}}
\partial_{\bar{\sigma}} X^{\mu}(\bar{\tau})  \frac{\partial}{\partial X^{\mu}(\bar{\tau})})
\tilde{H}^{\bot}_{d(\mu \bar{\sigma}')}
\label{zero}
\end{eqnarray}
 to \eqref{SecondOrderAction}, we can rewrite $H$ in \eqref{SecondOrderAction} as 
\begin{align}
&H(-i\frac{\partial}{\partial \bar{\tau}}, -i\frac{1}{\bar{e}}\frac{\partial}{\partial X(\bar{\tau})}, X(\bar{\tau}), \bar{h})
\nonumber \\
&\quad
=\frac{1}{2}\frac{1}{2\lambda\bar{\rho}}(-i\frac{\partial}{\partial \bar{\tau}})^2
\nonumber \\
&\qquad
+\int_0^{2\pi} d\bar{\sigma} \left( \sqrt{\bar{h}} 
\left(
\frac{1}{2}(-i\frac{1}{\bar{e}}\frac{\partial}{\partial X^{\mu}(\bar{\tau})})^2
+\frac{1}{2} \bar{e}^{-2} (\partial_{\bar{\sigma}}X^{\mu}(\bar{\tau}))^2
\right)
+i\bar{e} \bar{n}^{\bar{\sigma}} \partial_{\bar{\sigma}} X_{\mu}(\bar{\tau}) (-i\frac{1}{\bar{e}}\frac{\partial}{\partial X^{\mu}(\bar{\tau})})\right),
\nonumber \\ \label{bosonicHamiltonian}
\end{align}
where $\bar{n}^{\bar{\sigma}}(\bar{\sigma}, \bar{\tau})$ is the shift vector in the ADM formalism.

The propagator for $\tilde{H}^{\bot}_{d(\mu \bar{\sigma})}$ defined by
\begin{equation}
\Delta_F(\bar{h}, X(\bar{\tau}), \bar{\tau}; \; \bar{h},'  X'(\bar{\tau}'), \bar{\tau},')=<\tilde{H}^{\bot}_{d(\mu \bar{\sigma})} (\bar{h}, X(\bar{\tau}), \bar{\tau})
\tilde{H}^{\bot}_{d(\mu \bar{\sigma})}(\bar{h},'  X'(\bar{\tau}'), \bar{\tau}')>
\end{equation}
satisfies
\begin{equation}
H(-i\frac{\partial}{\partial \bar{\tau}}, -i\frac{1}{\bar{e}}\frac{\partial}{\partial X(\bar{\tau})}, X(\bar{\tau}), \bar{h})
\Delta_F(\bar{h}, X(\bar{\tau}), \bar{\tau}; \; \bar{h},'  X'(\bar{\tau}'), \bar{\tau},')
=
\delta(\bar{h}-\bar{h}') \delta(X(\bar{\tau})-X'(\bar{\tau}'))\delta(\bar{\tau}-\bar{\tau}'). \label{delta}
\end{equation}
In order to obtain a Schwinger representation of the propagator, we use the operator formalism $(\hat{\bar{h}}, \hat{X}(\hat{\bar{\tau}}), \hat{\bar{\tau}})$ of the first quantization, whereas the conjugate momentum is written as $(\hat{p}_{\bar{h}},  \hat{p}_{X}(\bar{\tau}), \hat{p}_{\bar{\tau}})$. The eigen state is given by $|\bar{h}, X(\bar{\tau}), \bar{\tau}>$. 

Since \eqref{delta} means that $\Delta_F$ is an inverse of $H$, $\Delta_F$ can be expressed by a matrix element of the operator $\hat{H}^{-1}$ as
\begin{equation}
\Delta_F(\bar{h}, X(\bar{\tau}), \bar{\tau}; \; \bar{h},'  X'(\bar{\tau}'), \bar{\tau},')
=
<\bar{h}, X(\bar{\tau}), \bar{\tau}| \hat{H}^{-1}(\hat{p}_{\bar{\tau}}, \hat{p}_{X}(\bar{\tau}), \hat{X}(\hat{\bar{\tau}}), \hat{\bar{h}}) |\bar{h},' X'(\bar{\tau}'), \bar{\tau}' >,
\label{InverseH}
\end{equation}
On the other hand,
\begin{eqnarray}
\hat{H}^{-1}=  \int _0^{\infty} dT e^{-T\hat{H}}, \label{IntegralFormula}
\end{eqnarray}
because
\begin{equation}
\lim_{\epsilon \to 0+} \int _0^{\infty} dT e^{-T(\hat{H}+\epsilon)}
=
\lim_{\epsilon \to 0+}  \left[\frac{1}{-(\hat{H}+\epsilon)} e^{-T(\hat{H}+\epsilon)}
\right]_0^{\infty}
=\hat{H}^{-1}.
\end{equation}
This fact and \eqref{InverseH} imply
\begin{equation}
\Delta_F(\bar{h}, X(\bar{\tau}), \bar{\tau}; \; \bar{h},'  X'(\bar{\tau}'), \bar{\tau}')
=
\int _0^{\infty} dT <\bar{h}, X(\bar{\tau}), \bar{\tau}|  e^{-T\hat{H}} |\bar{h},' X'(\bar{\tau}'), \bar{\tau}' >.
\end{equation}
In order to define two-point correlation functions that is invariant under the general coordinate transformations in the string geometry, we define in and out states as
\begin{eqnarray}
&&||X_i \,|\,h_f, ; h_i>_{in} \nonumber \\
&:=& \int_{h_i}^{h_f} \mathcal{D}h'|\bar{h},' X_i:=X'(\bar{\tau}'=-\infty), \bar{\tau}'=-\infty > \nonumber \\
&&<X_f\,|\,h_f, ; h_i||_{out} \nonumber \\
&:=& \int_{h_i}^{h_f} \mathcal{D} h <\bar{h}, X_f:=X(\bar{\tau}=\infty), \bar{\tau}=\infty|,
\end{eqnarray}
where $h_i$ and $h_f$ represent the metrics of the cylinders at $\bar{\tau}=\pm \infty$, respectively. When we insert asymptotic states, we integrate out $X_f$, $X_i$, $h_f$ and $h_i$ in the two-point correlation function for these states;
\begin{eqnarray}
\Delta_F(X_f; X_i|h_f, ; h_i) :=\int _0^{\infty} dT <X_f \,|\,h_f, ; h_i||_{out}  e^{-T\hat{H}} ||X_i \,|\,h_f, ; h_i>_{in}
\end{eqnarray}

In the same way as in \cite{Sato:2017qhj}, by inserting completeness relations of the eigen states, we obtain
\begin{align}
&\Delta_F(X_f; X_i|h_f, ; h_i) 
\nonumber \\
&\qquad
= \int_{h_i X_i, -\infty}^{h_f, X_f, \infty}  \mathcal{D} h \mathcal{D} X(\bar{\tau}) \mathcal{D}\bar{\tau} 
\int \mathcal{D} T  
\int 
\mathcal{D} p_T
\mathcal{D}p_{X} (\bar{\tau})
\mathcal{D}p_{\bar{\tau}}
 \nonumber \\
&\qquad\qquad
\exp \Biggl(- \int_{-\infty}^{\infty} dt \Bigr(
-i p_{T}(t) \frac{d}{dt} T(t) 
-i p_{\bar{\tau}}(t)\frac{d}{dt}\bar{\tau}(t)
-i p_{X}(\bar{\tau}, t)\cdot \frac{d}{dt} X(\bar{\tau}, t)\nonumber \\
&\qquad\qquad\qquad
+T(t) H(p_{\bar{\tau}}(t), p_{X}(\bar{\tau}, t), X(\bar{\tau}, t), \bar{h})\Bigr) \Biggr),  \label{canonicalform}
\end{align}
where $p_{X}(\bar{\tau}, t) \cdot \frac{d}{dt} X(\bar{\tau}, t):= \int d\bar{\sigma} \bar{e} p_{X}^{\mu}(\bar{\tau}, t) \frac{d}{dt} X_{\mu}(\bar{\tau}, t)$.

By integrate out $p_{\bar{\tau}}(t)$ and $p_{X}(\bar{\tau}, t)$ by using the relation of the ADM formalism, we obtain
\begin{align}
&\Delta_F(X_f; X_i|h_f ; h_i) \nonumber \\
&\quad
= \int_{h_i X_i, -\infty}^{h_f, X_f, \infty} 
\mathcal{D} T
\mathcal{D} h  \mathcal{D} X(\bar{\tau})\mathcal{D} \bar{\tau} 
\mathcal{D} p_T 
\nonumber \\
&\qquad\quad
\exp \Biggl(- \int_{-\infty}^{\infty} dt \Bigl(-i p_{T}(t) \frac{d}{dt} T(t)   +\lambda\bar{\rho}\frac{1}{T(t)}(\frac{d \bar{\tau}(t)}{dt})^2\nonumber \\
&\qquad\qquad\qquad
+\int d\bar{\sigma} \sqrt{\bar{h}} ( 
\frac{1}{2}\bar{h}^{00}\frac{1}{T(t)}\partial_{t} X^{\mu}(\bar{\sigma}, \bar{\tau}, t)\partial_{t} X_{\mu}(\bar{\sigma}, \bar{\tau}, t) +\bar{h}^{01}\partial_{t} X^{\mu}(\bar{\sigma}, \bar{\tau}, t)\partial_{\bar{\sigma}} X_{\mu}(\bar{\sigma}, \bar{\tau}, t) \nonumber \\
&\qquad\qquad\qquad
+\frac{1}{2}\bar{h}^{11}T(t)\partial_{\bar{\sigma}} X^{\mu}(\bar{\sigma}, \bar{\tau}, t)\partial_{\bar{\sigma}} X_{\mu}(\bar{\sigma}, \bar{\tau}, t)
) \Bigr) \Biggr). \label{pathint1}
\end{align}
We should note that the time derivative in \eqref{pathint1} is in terms of $t$, not $\bar{\tau}$ at this moment. In the following, we will see that $t$ can be fixed to $\bar{\tau}$ by using a reparametrization of $t$ that parametrizes a trajectory.

By inserting
$\int \mathcal{D}c \mathcal{D}b
e^{\int_0^{1} dt \left(\frac{d b(t)}{dt} \frac{d c(t)}{dt}\right)
},$
where $b(t)$ and $c(t)$ are $bc$-ghost, we obtain 
\begin{align}
&\Delta_F(X_f; X_i|h_f ; h_i) \nonumber \\
&\quad
=Z_0 \int_{h_i X_i, -\infty}^{h_f, X_f, \infty} 
\mathcal{D} T
\mathcal{D} h  \mathcal{D} X(\bar{\tau})\mathcal{D} \bar{\tau} 
\mathcal{D} p_T
\mathcal{D}c \mathcal{D}b  \nonumber \\
&\qquad\quad
\exp \Biggl(- \int_{-\infty}^{\infty} dt \Bigl(
-i p_{T}(t) \frac{d}{dt} T(t)  
 +\lambda\bar{\rho}\frac{1}{T(t)}(\frac{d \bar{\tau}(t)}{dt})^2 +\frac{d b(t)}{dt} \frac{d (T(t) c(t))}{dt}\nonumber \\
&\qquad\qquad\qquad
+\int d\bar{\sigma} \sqrt{\bar{h}} ( 
\frac{1}{2}\bar{h}^{00}\frac{1}{T(t)}\partial_{t} X^{\mu}(\bar{\sigma}, \bar{\tau}, t)\partial_{t} X_{\mu}(\bar{\sigma}, \bar{\tau}, t) +\bar{h}^{01}\partial_{t} X^{\mu}(\bar{\sigma}, \bar{\tau}, t)\partial_{\bar{\sigma}} X_{\mu}(\bar{\sigma}, \bar{\tau}, t) \nonumber \\
&\qquad\qquad\qquad
+\frac{1}{2}\bar{h}^{11}T(t)\partial_{\bar{\sigma}} X^{\mu}(\bar{\sigma}, \bar{\tau}, t)\partial_{\bar{\sigma}} X_{\mu}(\bar{\sigma}, \bar{\tau}, t)
) \Bigr) \Biggr), 
\label{PropWMult}
\end{align}
where we have redefined as $c(t) \to T(t) c(t)$, and $Z_0$ represents an overall constant factor. In the following, we will rename it $Z_1, Z_2, \cdots$ when the factor changes.

The integrand variable $p_T (t)$ plays the role of the Lagrange multiplier providing the following condition,
\begin{align}
F_1(t):=\frac{d}{dt}T(t)=0,
\label{F1gauge}
\end{align}
which can be understood as a gauge fixing condition. Indeed, by choosing this gauge in
\begin{align}
&\Delta_F(X_f; X_i|h_f ; h_i) \nonumber \\
&\quad
=Z_1 \int_{h_i X_i, -\infty}^{h_f, X_f, \infty} 
\mathcal{D} T
\mathcal{D} h  \mathcal{D} X(\bar{\tau})\mathcal{D} \bar{\tau} 
\nonumber \\
&\qquad\quad
\exp \Biggl(- \int_{-\infty}^{\infty} dt \Bigl(
\lambda\bar{\rho}\frac{1}{T(t)}(\frac{d \bar{\tau}(t)}{dt})^2 
+\int d\bar{\sigma} \sqrt{\bar{h}} ( 
\frac{1}{2}\bar{h}^{00}\frac{1}{T(t)}\partial_{t} X^{\mu}(\bar{\sigma}, \bar{\tau}, t)\partial_{t} X_{\mu}(\bar{\sigma}, \bar{\tau}, t)
\nonumber \\
&\qquad\qquad\qquad
 +\bar{h}^{01}\partial_{t} X^{\mu}(\bar{\sigma}, \bar{\tau}, t)\partial_{\bar{\sigma}} X_{\mu}(\bar{\sigma}, \bar{\tau}, t) 
+\frac{1}{2}\bar{h}^{11}T(t)\partial_{\bar{\sigma}} X^{\mu}(\bar{\sigma}, \bar{\tau}, t)\partial_{\bar{\sigma}} X_{\mu}(\bar{\sigma}, \bar{\tau}, t)
) \Bigr) \Biggr),
\label{pathint2}
\end{align}
we obtain \eqref{PropWMult}.
The expression \eqref{pathint2} has a manifest one-dimensional diffeomorphism symmetry with respect to $t$, where $T(t)$ is transformed as an einbein \cite{Schwinger0}. 

Under $\frac{d\bar{\tau}}{d\bar{\tau}'}=T(t)$, which implies
\begin{eqnarray}
\bar{h}^{00}&=&T^2\bar{h}^{'00} \nonumber \\
\bar{h}^{01}&=&T\bar{h}^{'01} \nonumber \\
\bar{h}^{11}&=&\bar{h}^{'11} \nonumber \\
\sqrt{\bar{h}}&=&\frac{1}{T}\sqrt{\bar{h}'} \nonumber \\
\bar{\rho}&=&\frac{1}{T}\bar{\rho}' \nonumber \\
X^{\mu}(\bar{\sigma}, \bar{\tau}, t)&=&X^{'\mu}(\bar{\sigma}, \bar{\tau}', t)
\nonumber \\
(\frac{d \bar{\tau}(t)}{dt})^2
&=&
T^2
(\frac{d \bar{\tau}'(t)}{dt})^2,
\end{eqnarray}

$T(t)$ disappears in \eqref{pathint2} and we obtain 
\begin{align}
&\Delta_F(X_f; X_i|h_f ; h_i) \nonumber \\
&=
Z_2 \int_{h_i X_i, -\infty}^{h_f, X_f, \infty} 
\mathcal{D} h  \mathcal{D} X(\bar{\tau})\mathcal{D} \bar{\tau} 
\nonumber \\
&\qquad
\exp \Biggl(- \int_{-\infty}^{\infty} dt \Bigl(
\lambda\bar{\rho}(\frac{d \bar{\tau}(t)}{dt})^2 \nonumber \\
&+\int d\bar{\sigma} \sqrt{\bar{h}} ( 
\frac{1}{2}\bar{h}^{00}\partial_{t} X^{\mu}(\bar{\sigma}, \bar{\tau}, t)\partial_{t} X_{\mu}(\bar{\sigma}, \bar{\tau}, t) +\bar{h}^{01}\partial_{t} X^{\mu}(\bar{\sigma}, \bar{\tau}, t)\partial_{\bar{\sigma}} X_{\mu}(\bar{\sigma}, \bar{\tau}, t) \nonumber \\
&+\frac{1}{2}\bar{h}^{11}\partial_{\bar{\sigma}} X^{\mu}(\bar{\sigma}, \bar{\tau}, t)\partial_{\bar{\sigma}} X_{\mu}(\bar{\sigma}, \bar{\tau}, t)
) \Bigr) \Biggr). \label{pathint3}
\end{align}
This action is still invariant under the diffeomorphism with respect to t if $\bar{\tau}$ transforms in the same way as $t$. 

\begin{figure}[htb]
\centering
\includegraphics[width=6cm]{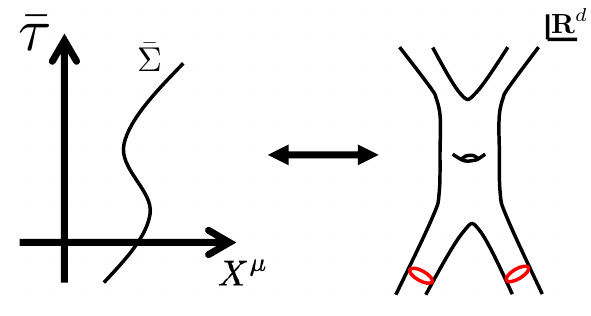}
\caption{A path and a Riemann surface. The line on the left is a trajectory in the path integral. The trajectory parametrized by $\bar{\tau}$ from $\bar{\tau}=-\infty$ to $\bar{\tau}=\infty$, represents a Riemann surface with fixed punctures in $\bold{R}^{d}$ on the right.} 
\label{Pathintegral}
\end{figure}

If we choose a different gauge
\begin{equation}
F_2(t):=\bar{\tau}-t=0, \label{F2gauge}
\end{equation} 
in \eqref{pathint3}, we obtain 
\begin{align}
&\Delta_F(X_f; X_i|h_f ; h_i) \nonumber \\
&\quad
=Z_3 \int_{h_i X_i, -\infty}^{h_f, X_f, \infty} 
\mathcal{D} h  \mathcal{D} X(\bar{\tau})\mathcal{D} \bar{\tau} 
\mathcal{D} \alpha \mathcal{D}c \mathcal{D}b
\nonumber \\
&\qquad\quad
\exp \Biggl(- \int_{-\infty}^{\infty} dt \Bigl( +\alpha(t) (\bar{\tau}-t) +b(t)c(t)(1-\frac{d \bar{\tau}(t)}{dt})  +\lambda\bar{\rho}(\frac{d \bar{\tau}(t)}{dt})^2 \nonumber \\
&\qquad\qquad\qquad
+\int d\bar{\sigma} \sqrt{\bar{h}} ( 
\frac{1}{2}\bar{h}^{00}\partial_{t} X^{\mu}(\bar{\sigma}, \bar{\tau}, t)\partial_{t} X_{\mu}(\bar{\sigma}, \bar{\tau}, t) +\bar{h}^{01}\partial_{t} X^{\mu}(\bar{\sigma}, \bar{\tau}, t)\partial_{\bar{\sigma}} X_{\mu}(\bar{\sigma}, \bar{\tau}, t) \nonumber \\
&\qquad\qquad\qquad\qquad\qquad\qquad
+\frac{1}{2}\bar{h}^{11}\partial_{\bar{\sigma}} X^{\mu}(\bar{\sigma}, \bar{\tau}, t)\partial_{\bar{\sigma}} X_{\mu}(\bar{\sigma}, \bar{\tau}, t)
) \Bigr) \Biggr) \nonumber \\
&\quad
=Z\int_{h_i, X_i}^{h_f, X_f} 
\mathcal{D} h  \mathcal{D} X
\nonumber \\
&\qquad\quad
\exp \Biggl(- \int_{-\infty}^{\infty} d\bar{\tau} 
\int d\bar{\sigma} \sqrt{\bar{h}} ( 
\frac{\lambda }{4\pi}\bar{R}(\bar{\sigma}, \bar{\tau})
+\frac{1}{2}\bar{h}^{00}\partial_{\bar{\tau}} X^{\mu}(\bar{\sigma}, \bar{\tau})\partial_{\bar{\tau}} X_{\mu}(\bar{\sigma}, \bar{\tau})
+\bar{h}^{01}\partial_{\bar{\tau}} X^{\mu}(\bar{\sigma}, \bar{\tau})\partial_{\bar{\sigma}} X_{\mu}(\bar{\sigma}, \bar{\tau}) 
 \nonumber \\
&\qquad\qquad\qquad\qquad\qquad\qquad\qquad
+\frac{1}{2}\bar{h}^{11}\partial_{\bar{\sigma}} X^{\mu}(\bar{\sigma}, \bar{\tau})\partial_{\bar{\sigma}} X_{\mu}(\bar{\sigma}, \bar{\tau})
) \Biggr). \label{prelastaction}
\end{align}
The path integral is defined over all possible two-dimensional Riemannian manifolds with fixed punctures in $\bold{R}^{d}$ as in Fig. \ref{Pathintegral}.
The diffeomorphism times Weyl invariance of the action in \eqref{prelastaction} implies that the correlation function is given by 
\begin{equation}
\Delta_F(X_f; X_i|h_f ; h_i)
=
Z
\int_{h_i, X_i}^{h_f, X_f} 
\mathcal{D} h  \mathcal{D} X
e^{-\lambda \chi}
e^{-S_{s}}, 
\label{FinalPropagator}
\end{equation}
where
\begin{equation}
S_{s}
=
\int_{-\infty}^{\infty} d\tau \int d\sigma \sqrt{h(\sigma, \tau)} \left(\frac{1}{2} h^{mn} (\sigma, \tau) \partial_m X^{\mu}(\sigma, \tau) \partial_n X_{\mu}(\sigma, \tau) \right),
\end{equation}
and $\chi$ is the Euler number of the two-dimensional Riemannian manifold.
For regularization, we divide the correlation function by $Z$ and the volume of the diffeomorphism and the Weyl transformation $V_{diff \times Weyl} $, by renormalizing $\tilde{H}^{\bot}_{d(\mu \bar{\sigma})}$.  \eqref{FinalPropagator} is the all-order  perturbative partition function of the string theory itself that possesses the moduli in the string theory. Especially, in string geometry, the consistency of the perturbation theory around the background \eqref{solution} determines $d=26$ (the critical dimension).

\section{Conclusion and Discussion}
\setcounter{equation}{0}
The Newtonian limit of any string geometry model around the flat background reproduces the perturbative string theory on the flat spacetime.  

In case of coupling with a $u(1)$ gauge field, the perturbative vacuum solution in \cite{Sato:2017qhj} and the Newtonian potential background in this paper coincide in the limit to reproduce the perturbative string. In this case, there exists accidentally an analytic solution that reduces to the Newtonian potential background in the limit, namely the perturbative vacuum solution. Thus, there does not exist two vacua to reproduce the perturbative string on the flat spacetime. 

In general, it is natural to expect that one can reproduce the perturbative string theory on any supergravity background if one takes the Newtonian limit of any string geometry model around the corresponding background that includes the supergravity background \cite{Honda-Sato}.

\section*{Acknowledgements}
We would like to thank  
M. Honda,
A. Tsuchiya
T. Yoneya
and especially 
H. Kawai for notifying us of the structure of the Newtonian limit when we derive the perturbative strings from the string geometry theory. 
YS was supported by a grant from the NSF of China with Grant No: 11947301.

\appendix

\section{The relation between the perturbative vacuum background and the Newtonian potential background}
\setcounter{equation}{0}
The perturbative string theory is reproduced from fluctuations around a perturbative vacuum background of the Einstein-Hilbert action in string geometry theory coupled with a $u(1)$ gauge field in \cite{Sato:2017qhj}. On the other hand,  the perturbative string theory is also reproduced from fluctuations around the Newtonian potential background of an arbitrary gravitational action in string geometry theory in this paper.  In this appendix, we clarify the relation between the perturbative vacuum background and the Newtonian potential background. 

The background that represents the perturbative vacuum  is given by
\begin{eqnarray}
\bar{ds}^2
&=& 2\lambda \bar{\rho}(\bar{h}) N^2(X(\bar{\tau})) (dX^d)^2 +\int d\bar{\sigma}   \bar{e} \int d\bar{\sigma}' \bar{e}' N^{\frac{2}{2-D}}(X(\bar{\tau})) \frac{\bar{e}^3(\bar{\sigma}, \bar{\tau})}{\sqrt{\bar{h}(\bar{\sigma}, \bar{\tau})}} \delta_{(\mu \bar{\sigma}) (\mu' \bar{\sigma}')}
d X^{(\mu \bar{\sigma})} d X^{(\mu' \bar{\sigma}')}, \nonumber \\
\bar{A}_d&=&i \sqrt{\frac{2-2D}{2-D}}\frac{\sqrt{2\lambda \bar{\rho}(\bar{h}) }}{\sqrt{G_N}} N(X(\bar{\tau})), \qquad
\bar{A}_{(\mu \bar{\sigma})}=0, \label{OldSolution}
\end{eqnarray}
where  $N(X(\bar{\tau}))=\frac{1}{1-\frac{1}{2}\alpha \phi(X(\bar{\tau}))}$. This is a solution to the equations of motion,
\begin{eqnarray}
&&R_{IJ}=G_N(\frac{1}{2} F_{IK} F_J^K +\frac{1}{2(2-D)}G_{IJ}|F|^2),  \label{EinsteinEq} \\
&&\nabla_K F^{KI}=0.
\end{eqnarray}
By expanding this background metric up to the first order in $\alpha$, we obtain the Newtonian potential background \eqref{solution}.  The contribution from the solution to the right-hand side of \eqref{EinsteinEq} becomes 0 in the first order because the contribution from the gauge field starts from the second order. Then, the solution satisfies \eqref{RicciFlat}. 
 In section 3, we find that the part of the action in the second order in the fluctuations \eqref{SecondOrderAction} coincides with  the one obtained in \cite{Sato:2017qhj} by rescaling an overall factor of the potential \eqref{DefPhi}, which is a solution of \eqref{diffeq}, although there are contributions from the fluctuations around the background of the $u(1)$ gauge field. 

Therefore, the perturbative string theory is reproduced in the case of the string geometry model coupled with the $u(1)$ gauge field  by the same mechanism  both in \cite{Sato:2017qhj}  and in this paper.  Thus, the Newtonian limit is a natural generalization of  reproducing the perturbative string theory in \cite{Sato:2017qhj} to the case of any string geometry model.

\section{Explicit form of the Lagrangian up to the second order in the fluctuations.}\label{DerivExpResult}
Here we provide the explicit form of (\ref{fluctuationAction}):
\begin{equation}
\scalebox{0.7}{$\displaystyle
\begin{aligned}
&L_0=
\frac{\alpha^2}{G_N} \frac{3D^2-9D+10}{4(D-2)^2} \bar{\alpha}_d^{-2} \bar{\alpha}_i^{-1} \partial^i \phi \partial_i \phi,
\\ 
&L_1=
\frac{1}{\sqrt{G_N}} \biggl\{ -\frac{(D-1)(D-6)}{8(D-2)^2}\alpha^2 \bar{\alpha}_d^{-2} \bar{\alpha}_i^{-1}  \partial^i \phi \partial_i \phi \tilde{H}_{dd}  
+\frac{3(D-1)}{8(D-2)} \alpha^2  \bar{\alpha}_d^{-2} \bar{\alpha}_i^{-1}  \partial^i \phi \partial_i \phi \tilde{H}^j_{j}
  \\
&\qquad\qquad\qquad
+ \biggl(
- \frac{D-1}{4(D-2)} \alpha^2 \bar{\alpha}_d^{-1}  \partial_i  \phi \partial_j \phi 
+ \frac{D-1}{4(D-2)} \alpha^2 \bar{\alpha}_d^{-1} \phi \partial_i \partial_j \phi  \biggr) \bar{\alpha}_d^{-1} \bar{\alpha}_i^{-\frac{1}{2}} \bar{\alpha}_j^{-\frac{1}{2}}   \tilde{H}^{ij}
\biggr\},
\\ 
&L_{2}=
\bar{\alpha}_d^{-1}
 \tilde{H}_{dd} 
\biggl\{
-\frac{1}{8} \left(1 +\frac{3D-5}{D-2} \alpha \bar{\alpha}_d^{-1} \phi + \frac{24D^2-85D+76}{4(D-2)^2} \alpha^2 \bar{\alpha}_d^{-2}\phi^2  \right)  \partial_d^2 
-\frac{1}{8} \left(1 +2 \alpha \bar{\alpha}_d^{-1}\phi + \frac{12D^2-49D+52}{4(D-2)^2} \alpha^2 \bar{\alpha}_d^{-2} \phi^2  \right)  \bar{\alpha}_i^{-1} \partial^i \partial_i 
 \\
&\qquad\qquad\qquad
+\frac{1}{4}\left(1+\frac{11D-23}{4(D-2)}\alpha \bar{\alpha}_d^{-1} \phi \right)\alpha  \bar{\alpha}_i^{-1}  \partial_i \phi   \partial^i 
-\frac{D^2-7D+6}{32(D-2)^2}  \alpha^2 \bar{\alpha}_d^{-1} \bar{\alpha}_i^{-1} \partial^i \phi \partial_i  \phi 
 \biggr\}
 \tilde{H}_{dd} 
 \\
 &\quad\quad
+ \bar{\alpha}_d^{-1}
\tilde{H}_{di} 
\biggl\{
-\frac{1}{2}\left(1 +2 \alpha \bar{\alpha}_d^{-1}\phi + \frac{12D^2- 49D+52}{4(D-2)^2} \alpha^2 \bar{\alpha}_d^{-2} \phi^2  \right)
 \bar{\alpha}_i^{-1}
 \delta^{ij}  \partial_d^2
-\frac{1}{2}\left(1 +\frac{D-3}{D-2} \alpha \bar{\alpha}_d^{-1}\phi +\frac{4D^2-21D+32}{4(D-2)^2} \alpha^2 \bar{\alpha}_d^{-2} \phi^2  \right) \bar{\alpha}_k^{-1}
  \delta^{ij} \partial^k \partial_k
  \\
&\qquad\qquad\qquad\quad
+\frac{1}{4(D-2)}\left(1 +\frac{D-4}{D-2} \alpha \bar{\alpha}_d^{-1} \phi \right) \alpha  \bar{\alpha}_i^{-\frac{1}{2}} \alpha_j^{-\frac{1}{2}}  \partial^i \phi \partial^j   
-\frac{1}{2(D-2)}\left(1+\frac{D-4}{D-2} \alpha \bar{\alpha}_d^{-1} \phi \right) \alpha  \bar{\alpha}_i^{-\frac{1}{2}} \alpha_j^{-\frac{1}{2}}  \partial^j \phi  \partial^i 
  \\
&\qquad\qquad\qquad\quad
-\frac{1}{2(D-2)}\left((D-3) +\frac{3D^2-17D+26}{2(D-2)} \alpha \bar{\alpha}_d^{-1} \phi \right) \alpha \bar{\alpha}_k^{-1}   \partial_k \phi \delta^{ij} \partial_k 
-\frac{D-9}{8(D-2)}  \alpha^2 \bar{\alpha}_d^{-1} \delta^{ij} \bar{\alpha}_k^{-1} \partial^k \phi \partial_k  \phi 
+\frac{D-3}{2(D-2)}  \alpha^2 \bar{\alpha}_d^{-1} \bar{\alpha}_i^{-\frac{1}{2}}  \bar{\alpha}_j^{-\frac{1}{2}} \partial_i \phi \partial_j  \phi 
\biggr\} 
\tilde{H}_{dj} 
  \\
 &\qquad\quad
+\tilde{H}_{dd}
\biggl\{
\frac{1}{4}\left(1+ 2\alpha \bar{\alpha}_d^{-1}\phi + \frac{12D^2-49D+52}{4(D-2)^2} \alpha^2 \bar{\alpha}_d^{-2} \phi^2  \right) \bar{\alpha}_d^{-1}  \delta^{ij}   \partial_d^2
+\frac{1}{4}\left(1 +2\alpha \bar{\alpha}_d^{-1}\phi + \frac{4D^2-21D+32}{4(D-2)^2} \alpha^2 \bar{\alpha}_d^{-2} \phi^2  \right) \bar{\alpha}_d^{-1} \bar{\alpha}_k^{-1}
\bar{\alpha}_d \bar{\alpha}_j \delta^{ij}  \partial^k \partial_k 
 \\
&\qquad\qquad\qquad\quad
-\frac{1}{8(D-2)}\left((3D-5)-\frac{10D^2-55D-84}{2(D-2)} \alpha \bar{\alpha}_d^{-1} \phi \right) \alpha \bar{\alpha}_d^{-1} \bar{\alpha}_k^{-1} \delta^{ij}   \partial_k \phi  \partial^k 
-\frac{1}{2}\left(1 +\frac{3(D-3)}{2(D-2)}\alpha \bar{\alpha}_d^{-1} \phi \right) \alpha \bar{\alpha}_d^{-1} \bar{\alpha}_i^{-\frac{1}{2}} \bar{\alpha}_i^{-\frac{1}{2}}  \phi \partial_j \phi  \partial_i 
\phi  \partial_j \phi    
\\
&\qquad\qquad\qquad\quad
-\frac{1}{2}\left(1 +\frac{3(D-3)}{2(D-2)}\alpha \bar{\alpha}_d^{-1} \phi \right) \alpha \bar{\alpha}_d^{-1} \bar{\alpha}_i^{-\frac{1}{2}} \bar{\alpha}_i^{-\frac{1}{2}} \overset{\leftarrow}{\partial_i} 
+\frac{5D^2-27D+38}{16(D-2)^2} \alpha^2 \bar{\alpha}_d^{-2}  \bar{\alpha}_j  \bar{\alpha}_k^{-1}   \delta^{ij}  \partial^k \phi \partial_k \phi 
+\frac{3D-7}{8(D-2)}  \alpha^2 \bar{\alpha}_d^{-2} \bar{\alpha}_i^{-\frac{1}{2}}  \bar{\alpha}_j^{-\frac{1}{2}} \partial_i \phi \partial_j  \phi 
\biggr\} 
 \tilde{H}_{ij} 
\\
&\qquad\quad
+ \tilde{H}_{ij} 
\biggl\{
\frac{1}{8}\left(1- \frac{D-3}{D-2} \alpha \bar{\alpha}_d^{-1}\phi + \frac{4D^2 -21D+32}{4(D-2)^2} \alpha^2 \bar{\alpha}_d^{-2} \phi^2  \right) \bar{\alpha}_d^{-1} 
\biggl(
-2 \delta^{ik}\delta^{jl}  \partial_d^2 + \delta^{ij}\delta^{kl} \partial_d^2 
\biggr)
 \\
&\qquad\qquad\qquad\quad
-\frac{1}{8(D-2)}\left(1 +\frac{D^2-3D-10}{2(D-2)} \alpha \bar{\alpha}_d^{-1} \phi \right) \alpha \bar{\alpha}_d^{-1} \bar{\alpha}_m^{-1}  \partial_m \phi
\biggl( 
-  2 \delta^{ik}\delta^{jl}  \partial^m + \delta^{ij}\delta^{kl} \partial^m
\biggr)
 \\
&\qquad\qquad\qquad\quad
+\frac{1}{8}\left( 1- \frac{3}{D-2} \alpha \bar{\alpha}_d^{-1}\phi + \frac{6}{(D-2)^2} \alpha^2 \bar{\alpha}_d^{-2} \phi^2  \right)  \bar{\alpha}_m^{-1}
\biggl(
- 2 \delta^{ik}\delta^{jl} \partial^m  \partial_m +  \delta^{ij}\delta^{kl} \partial^m \partial_m 
\biggr)
 \\
&\qquad\qquad\qquad\quad
+\frac{3D^2-9D-2}{16(D-2)^2}  \alpha^2 \bar{\alpha}_d^{-2} \bar{\alpha}_m^{-1} \partial^m \phi \partial_m  \phi  \delta^{ik}\delta^{jl} 
-\frac{3^2-9D-10}{32(D-2)^2}  \alpha^2 \bar{\alpha}_d^{-2} \bar{\alpha}_m^{-1} \partial^m \phi \partial_m  \phi  \delta^{ij}\delta^{kl} 
+\frac{D^2-3D-4}{4(D-2)^2} \alpha^2  \alpha \bar{\alpha}_d^{-2} \bar{\alpha}_i^{-\frac{1}{2}} \bar{\alpha}_j^{-\frac{1}{2}}  \partial_i \phi \partial_k  \phi \delta^{jl}
 \\
&\qquad\qquad\qquad\quad
+\frac{D^2-3D-4}{4(D-2)^2}\alpha^2 \bar{\alpha}_d^{-2}  \bar{\alpha}_i^{-\frac{1}{2}} \bar{\alpha}_j^{-\frac{1}{2}}  \partial_i \phi \partial_j \phi  \delta^{kl} 
-\frac{1}{D-2}\left(1+\frac{D^2-3D-10}{4(D-2)} \alpha \bar{\alpha}_d^{-1} \phi \right) \alpha \bar{\alpha}_d^{-1} \bar{\alpha}_i^{-\frac{1}{2}} \alpha_k^{-\frac{1}{2}} \partial_i \phi \delta^{jl} \partial^k 
 \\
&\qquad\qquad\qquad\quad
+\frac{1}{2(D-2)}\left(1-\frac{D^2-3D-10}{4(D-2)} \alpha \bar{\alpha}_d^{-1} \phi \right) \alpha \bar{\alpha}_d^{-1} \bar{\alpha}_i^{-\frac{1}{2}} \bar{\alpha}_k^{-\frac{1}{2}}  \partial_k  \phi \delta^{jl} \partial_i
+\frac{D^2+3D-14}{4(D-2)^2} \alpha^2 \bar{\alpha}_d^{-2} \bar{\alpha}_i^{-\frac{1}{2}} \alpha_k^{-\frac{1}{2}} \partial_i \phi \partial_k \phi \delta^{jl}
 \\
&\qquad\qquad\qquad\quad
-\frac{1}{2(D-2)}\left(1-\frac{D^2-3D-4}{2(D-2)} \alpha \bar{\alpha}_d^{-1} \phi \right) \alpha \bar{\alpha}_d^{-1} \bar{\alpha}_k^{-\frac{1}{2}} \bar{\alpha}_l^{-\frac{1}{2}}  \partial_l  \phi \delta^{ij}  \partial_k
-\frac{1}{2(D-2)}\left(1-\frac{D^2-3D-4}{2(D-2)} \alpha \bar{\alpha}_d^{-1} \phi \right) \alpha \bar{\alpha}_d^{-1} \bar{\alpha}_i^{-\frac{1}{2}} \bar{\alpha}_j^{-\frac{1}{2}}   \partial_j  \phi \delta^{ij} \partial_i
\biggr\}  \tilde{H}_{kl}  
\\
&\qquad\quad
+\tilde{H}_{dd}
\biggl\{ 
\left(1+3\alpha \bar{\alpha}_d^{-1} \phi \right)\alpha \bar{\alpha}_d^{-\frac{3}{2}} \bar{\alpha}_i^{-\frac{1}{2}} \partial_i \phi
\biggr\}
\partial_d \tilde{H}_{d}^i   
+\tilde{H}_{dj}
\biggl\{
\frac{1}{2}\left(1 +\frac{2(2D-5)}{D-2} \alpha \bar{\alpha}_d^{-1} \phi \right) \alpha \bar{\alpha}_d^{-\frac{3}{2}} \bar{\alpha}_i^{-\frac{1}{2}}  \partial_i \phi
\biggr\}
 \partial_d \tilde{H}^{ij},  
 \end{aligned}
 $}
\end{equation}
where we use short notation: we use indices $i$ instead of $(\mu \bar{\sigma})$ and all the indices are summed up. For an example, 
\begin{align}
\bar{\alpha}^{-1}_i \partial_i \phi \partial^i \tilde{H}_{dd}
={\rm i}\int d \bar{\sigma} \bar{e} \times \frac{\sqrt{\bar{h}}}{\bar{e}^3} \times \left( \sum_{\mu=0}^{d-1} \partial_{\bar{\sigma}}X_\mu (\bar{\tau}) \partial_{X^\mu (\bar{\tau})} \tilde{H}_{dd} \right).
\end{align}

\end{document}